\documentclass[A4paper,unsortedaddress,superscriptaddress,prl,preprint,showpacs]{revtex4-1}

\usepackage{hyperref}
\usepackage{epsfig}
\usepackage{graphicx}
\usepackage{subfigure}
\usepackage{latexsym}
\usepackage{color}
\usepackage{fullpage}
\usepackage{dcolumn}
\usepackage{bm}
\usepackage{ulem}
\usepackage{units}

\usepackage{units}

\begin{document}

\title{Strains Induced by Point Defects in Graphene on a Metal}%

\author{Nils Blanc}
\affiliation{Institut N\'{E}EL, CNRS \& Universit\'{e} Joseph Fourier -- BP166 -- F-38042 Grenoble Cedex 9 -- France}
\affiliation{CEA-UJF, INAC, SP2M, 17 rue des Martyrs, 38054 Grenoble Cedex 9 -- France}
\author{Fabien Jean}
\affiliation{Institut N\'{E}EL, CNRS \& Universit\'{e} Joseph Fourier -- BP166 -- F-38042 Grenoble Cedex 9 -- France}
\author{Arkady V. Krasheninnikov}
\affiliation{Department of Applied  Physics, Aalto University,  P.O. Box 11100, FI-00076 -- Finland \& Department of Physics, University of Helsinki, P.O. Box 43, FI-00014 -- Finland}
\author{Gilles Renaud}
\affiliation{CEA-UJF, INAC, SP2M, 17 rue des Martyrs, 38054 Grenoble Cedex 9 -- France}
\author{Johann Coraux}
\email{johann.coraux@grenoble.cnrs.fr}
\affiliation{Institut N\'{E}EL, CNRS \& Universit\'{e} Joseph Fourier -- BP166 -- F-38042 Grenoble Cedex 9 -- France}

\date{\today}%

\begin{abstract}

Strains strongly affect the properties of low-dimensional materials, such as graphene. By combining \textit{in situ}, \textit{in operando}, reflection high energy electron diffraction experiments with first-principles calculations, we show that large strains, above 2\%, are present in graphene during its growth by chemical vapor deposition on Ir(111) and when it is subjected to oxygen etching and ion bombardment. Our results unravel the microscopic relationship between point defects and strains in epitaxial graphene and suggest new avenues for graphene nanostructuring and engineering its properties through introduction of defects and intercalation of atoms and molecules between graphene and its metal substrate.

\end{abstract}

\pacs{81.05.ue, 61.72.J-, 81.15.Gh, 07.10.Pz, 68.60.Bs}

\maketitle

Strain engineering, sometimes referred to as "straintronics", is a powerful method for tuning the properties of bulk and two-dimensional materials \cite{Novoselov12}. Graphene, as an atomically-thin membrane with unprecedented mechanical strength \cite{Lee2008}, offers unique possibilities in this respect. While large compressive stresses applied to this material are readily relieved by wrinkle \cite{Bao2009} and nanobubble \cite{Levy2010} formation, tensile or small in-plane compressive strains can be stabilized by interaction with a substrate. Strain-induced changes in the vibrational properties  \cite{Mohiuddin2009,Ding2010}, electronic band-gaps \cite{Guinea2009,Levy2010}, as well as variations of the local \cite{Teague2009} and macroscopic electronic conductivity \cite{Huang2011,Fu2011}, have been reported. Strains are also expected around defects in graphene \cite{Cretu10prl}, and their role should be considered wherever defects are involved, for instance when engineering electronic \cite{Zhao2011} or magnetic \cite{Nair2012} properties using plasma etching or irradiation. Interactions between point defects have been predicted to be attractive or repulsive depending on their relative orientation and separation \cite{Kotakoski2006}, which could lead to the buildup of extended strains from local ones.

The experimental exploration of the interplay between defects in graphene and strains has started recently, with atmospheric condition Raman spectroscopy of ion-bombarded exfoliated graphene \cite{MartinsFerreira2010}. A step forward are \textit{in situ} investigations, which we report in this Letter, while the defects are created (or healed) under ultra-clean conditions (ultra-high vacuum). This minimizes the effects of contamination of graphene with highly reactive defects by molecular species present in air.

\begin{figure}[hbt]
  \begin{center}
  \includegraphics[width=80mm]{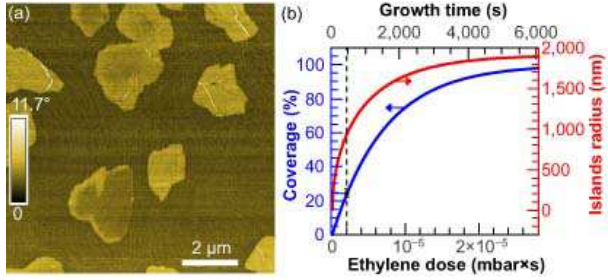}
  \caption{\label{fig1}(Color online) (a) AFM phase image of graphene islands (24\% coverage) grown on Ir(111) at 850$^\circ$C with ethylene. (b) Graphene coverage and average island radius (mainly relevant before $\simeq$ 25\%, when coalescence starts) as a function of ethylene dose and growth time. The vertical line corresponds to the ethylene dose for (a).}
  \end{center}
\end{figure}

Here, we report on the determination of the average strain, with a relative accuracy better than 10$^{-3}$, during the growth of graphene on Ir(111), its bombardment with ions and its etching with oxygen, using \textit{in situ}, \textit{in operando} reflection high energy electron diffraction (RHEED). By combining the observations with \textit{ab initio} density functional theory (DFT) calculations, we show that the global tensile strains stem from local ones around point defects which are formed during O$_2$ etching at high temperature and ion bombardment, or healed during chemical vapor deposition (CVD). Such strains are found to influence the epitaxy between graphene and the metal and can be used for nanostructuring epitaxial graphene and engineering its properties. 

CVD of graphene on Ir(111) under ultra-high vacuum with ethylene as a carbon precursor \cite{SupMat} at 850$^\circ$C started with a 9$\times$10$^{-2}$~$\mu$m$^{-2}$ density of nucleation centers, as determined \textit{ex situ} by atomic force microscopy (AFM) under ambient conditions [Fig.~\ref{fig1}(a)] \cite{SupMat}. The determination of graphene coverage based on a set of AFM images allowed one to calibrate ethylene dose. Indeed, coverage increases with dose following a modified Langmuir model without any free parameter \cite{Coraux2009}. In Fig.~\ref{fig1}(b), we show the average coverage of the sample, 24$\pm$5\%, which allows deducing the ethylene dose by using this model. From the average graphene island density, the average island radius is estimated to be $\sim$ 1~$\mu$m for the sample grown at 850$^\circ$C, assuming evenly-sized, disk-shaped islands.

\begin{figure}[hbt]
  \begin{center}
  \includegraphics[width=70mm]{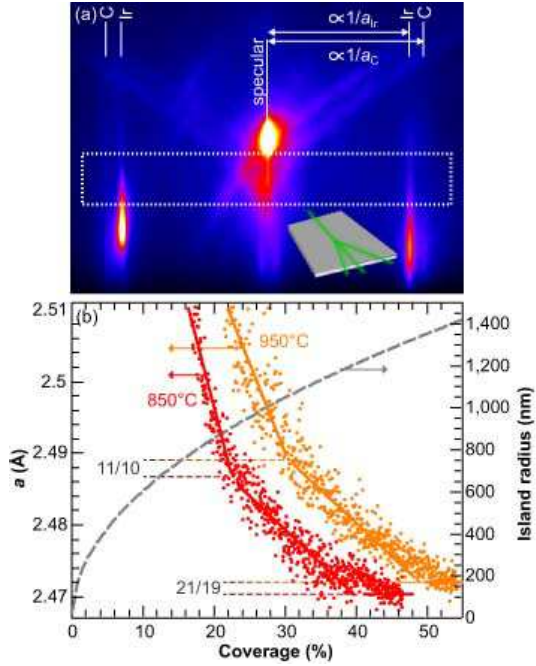}
  \caption{\label{fig2}(Color online) (a) RHEED pattern measured with 10~keV energy-electrons along the [$\bar{2}$11] direction of Ir(111), for a full graphene layer grown at 850$^\circ$C on Ir(111). The distances from the central streak to the other streaks are inversely proportional to the lattice parameters of graphene ($a_\mathrm{C}$) and Ir ($a_\mathrm{Ir}$) \cite{SupMat}. The dotted line-frame is the region used for determining the position of the various streaks as a function of ethylene dose. (b) Lattice parameter of graphene as a function of graphene coverage, for two growth temperatures. (colored lines are guides for the eyes), and average radius of the graphene islands for the sample grown at 850$^\circ$C (mainly relevant before $\simeq$ 25 \% when coalescence starts).}
  \end{center}
\end{figure}

The growth of graphene was studied with RHEED in real time at two growth temperatures. A typical RHEED pattern is shown in Fig.~\ref{fig2}(a) \cite{SupMat}. As expected for a flat crystalline surface, it displays streaks perpendicular to the surface. Two groups of streaks are visible on each side of the specularly reflected beam. Each comprises two streaks. The outer one only appears during graphene growth, and it is roughly 10\% farther away from the center of the pattern than the inner one. This streak is ascribed to graphene, whose lattice parameter is $\sim$ 10\% smaller (the distance from the corresponding streak to the center of reciprocal space is thus 10\% larger) than the Ir(111) one. We measured the distance between Ir and graphene peaks as a function of ethylene dose \cite{SupMat}. The distance between the Ir streaks served as a reference for lattice parameters, which are tabulated as a function of temperature \cite{Wimber1976}. Figure~\ref{fig2}(b) shows an overall decrease of about 1.6\% of $a_\mathrm{C}$, the surface projection of the lattice parameter in graphene, until steady values, of $a_\mathrm{C}$ = 2.4705$\pm$0.0020~\AA\, at 850$^\circ$C and $a_\mathrm{C}$ = 2.4723$\pm$0.0020~\AA\, at 950$^\circ$C, are reached, for graphene coverages above 50\% \cite{noteonTEC}. These values are about 0.6--0.7\% larger than those calculated for free-standing graphene \cite{Zakharchenko2009}, suggesting the presence of a residual tensile strain even in (almost) defect-free graphene. The $a_\mathrm{Ir}$ surface lattice parameter of Ir(111) (2.7319 and 2.7343~\AA\, at 850 and 950$^\circ$C) is a fractional number (21/19) times $a_\mathrm{C}$, \textit{i.e.} graphene and Ir(111) are commensurate. This implies a local periodic C-Ir interaction, giving rise to an energy gain presumably overcoming the cost associated with the tensile strain energy. Note that no significant increase in $a_\mathrm{C}$ is expected as arising from stress relief at the island edges \cite{Massies1993} because of the large size ($>$ 100~nm) of the islands.

\begin{figure*}[hbt]
  \begin{center}
  \includegraphics[width=135mm]{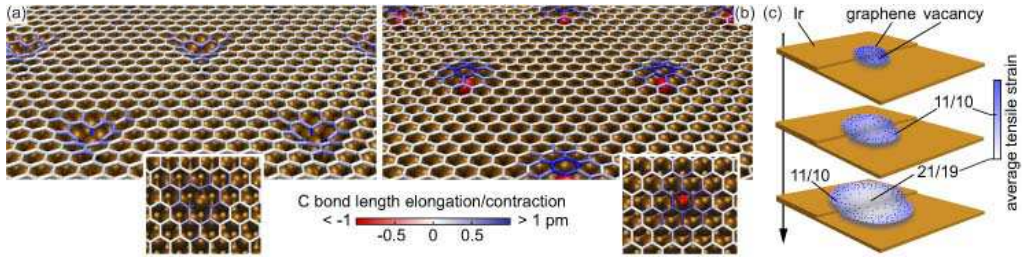}
  \caption{\label{fig3}(Color online) Structure of graphene on Ir(111) in the presence of  (a) single vacancies and (b) intercalated Ar atoms. Ir and Ar atoms are represented with ochre and red spheres respectively; C-C bonds are colored according to their length with regard to bonds in pristine graphene (see color bar). Insets are top-views. (c) Sketch of tensile strain (blue shades) in growing graphene islands resulting from the tensile strain due to vacancies (dark dots). Salient points match 11/10 and 21/19 commensurate structures (see text).}
  \end{center}
\end{figure*}

We now discuss the possible origins of the observed decrease in $a_\mathrm{C}$ with graphene coverage. The reduction of the graphene-metal interaction as the island size increases due to the decreasing contribution of edge atoms \cite{Lacovig2009}, which would relieve heteroepitaxial stress in graphene, can only be marginal given the fraction of edge atoms in the large islands considered here. The coalescence of neighboring islands, having different registries on Ir(111), implies the accommodation of one substrate interatomic distance over the distance between islands nucleation centers, $\sim$ 1~$\mu$m, \textit{i.e.} a negligible $\sim$ 0.03\% strain. More relevant are the numerous vacancies of various sizes that are trapped inside graphene at the growth front. At a 30\% coverage, their density is several 0.1~nm$^{-2}$ \cite{Coraux2009}, and $a_\mathrm{C}$ is several 0.1\% larger than the value at the end of growth. Our DFT calculations \cite{SupMat} for a defect density of 0.2~nm$^{-2}$ reveal that single-, di-, and tetra-vacancies in graphene/Ir(111) are surrounded by a tensile strain field [Fig.~\ref{fig3}(a) and Fig.~S3 of Supplemental Material \cite{SupMat}], from a few to several 0.1\% depending on the configuration, usually longer-ranged for larger vacancies (unless their location allows a close-to-perfect match between the positions of C dangling bonds and Ir atoms). These values are different from those expected in free-standing graphene \cite{Krasheninnikov2011} due to the strong interaction between C and metal atoms at vacancy edges \cite{Ugeda2011} [Fig.~\ref{fig3}(a)]. Even though this interaction reduces formation energies of vacancies \cite{Wang2013}, their migration barriers are high (3-8~eV, depending on the position in the moir\'{e} pattern and size of the vacancy), so that the agglomeration of vacancies, a situation reported to be energetically favorable for other types of defects in graphene \cite{Nguyen2012}, is hindered, especially for large defects like tetra-vacancies. Calculations of RHEED profiles from the atomic position optimized with DFT calculations qualitatively confirm the relevance of vacancy-induced strains (Fig.~S4 of Supplemental Material \cite{SupMat}).

The progressive filling of vacancies during growth, and the thermally-activated diffusion of small vacancies that are annihilated upon reaching the edges of graphene, are expected to decrease tensile strains, and thus to account for the decrease of $a_\mathrm{C}$. Due to the short lifetime of ethylene on graphene at the growth temperature, the filling of the vacancies must be less and less efficient as their size decreases, which agrees with the slower decrease in $a_\mathrm{C}$ at larger doses.

Moreover, the decrease in $a_\mathrm{C}$ is accompanied by a series of surface phase transitions: it shows salient points at 23 and 31\% coverage for 850 and 950$^\circ$C growth temperatures respectively, at $a_\mathrm{C}$=2.487 and 2.489~\AA, corresponding to a commensurate phase with 11 C rings on 10 Ir atoms (first order commensurability). Eventually, at 850 and 950$^\circ$C, $a_\mathrm{C}$ reaches 2.4705 and 2.4723~\AA\, steady values, both corresponding to 21 C rings on 19 Ir atoms, a superstructure corresponding to a second order commensurability similar to the one reported at room temperature in graphene/Ru(0001) \cite{Martoccia2008}. The change of the slope of $a_\mathrm{C}$ \textit{vs} dose points to a tendency of graphene to adopt the 11/10 phase, presumably because it maximizes the interaction between C and Ir. This implies, before the salient point, the coexistence of a 11/10 central region with a more strongly strained region around, where the vacancy density is higher [Fig.~\ref{fig3}(c)]. After the salient point the opposite situation is expected, with a decreasing vacancy density at the center of the graphene island eventually leading to a 21/19 phase, and a 11/10 region around the center of the island [Fig.~\ref{fig3}(c)]. How the slopes of $a_\mathrm{C}$ \textit{vs} dose change with temperature is a trade-off between vacancy diffusion, healing, and incorporation at edges, thus a complex balance between vacancy diffusion length (larger at higher temperature), inter-vacancy distance (likewise, larger at higher temperature), and graphene island radius (whose first time derivative, the growth rate of the islands, is smaller at higher temperature). An additional mechanism can be invoked: the first salient point coincides with the onset of coalescence, above which the graphene free edge length rapidly decreases. Micrometer-scale slippage of graphene on Ir(111) \cite{NDiaye2009}, which would promote the formation of a strained 11/10 phase stabilized by periodic C-Ir interactions, would then be hindered above this point and the slope of the $a_\mathrm{C}$ \textit{vs} dose would decrease.

\begin{figure*}[hbt]
  \begin{center}
  \includegraphics[width=150mm]{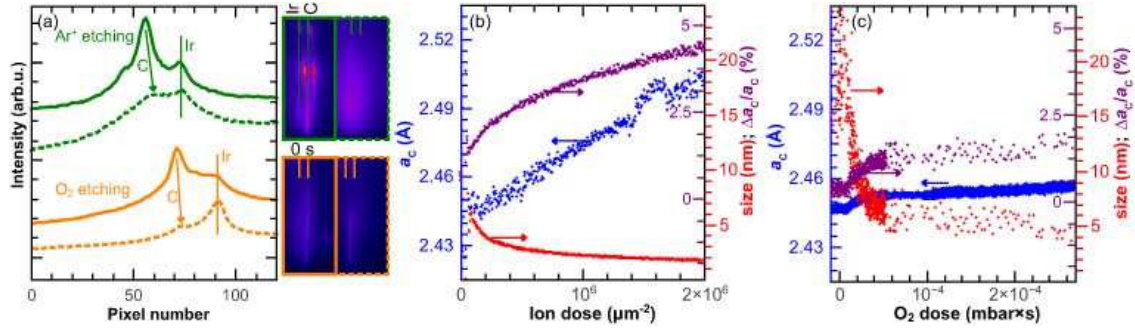}
  \caption{\label{fig4}(Color online) (a) RHEED pattern profiles at room temperature before and after Ar$^+$ bombardment with 200~eV ions (solid and doted green lines respectively), and at 950$^\circ$C before and at the end of O$_2$ etching (solid and doted orange lines respectively). RHEED patterns close to the first-order graphene and Ir streaks are shown aside before (0~s) and after O$_2$ etching and Ar$^+$ bombardment. (b,c) $a_\mathrm{C}$, size of structurally relevant domains, and relative distribution $\Delta a_\mathrm{C}/a_\mathrm{C}$ of lattice parameter as a function of ion (b) and O$_2$ (c) dose (see text for a discussion about the relevance of the $\Delta a_\mathrm{C}/a_\mathrm{C}$ and size estimates).}
  \end{center}
\end{figure*}

We now consider two other processes which are known to induce defects in graphene: ion bombardment and etching with oxygen. Once grown at 950$^\circ$C and cooled down to room temperature, graphene covering $\sim$ 100\% of the surface was bombarded with 200~eV Ar$^+$ ions. Such ions are expected to yield prominently single atom vacancies and Ar$^+$ ions trapped below graphene (referred to as interstitials in the following), with a yield close to unity \cite{Hahn1999}, and to leave Ir(111) essentially non-damaged. The average distance between defects, from a few nanometers to below one nanometer in the ion dose range explored, is smaller than between defects at initial stages of graphene growth. During bombardment, the graphene streaks shift towards the center of reciprocal space, broaden, and loose intensity [Fig.~\ref{fig4}(a)]. For an ion dose of about 2$\times$10$^6$~ions/$\mu$m$^2$, $a_\mathrm{C}$ is increased by as much as 2.2\% and the full-width at half maximum (FWHM) of the graphene streaks is multiplied by two before they vanish. Ir streaks only loose intensity but do not move or broaden. This intensity decrease upon increase of the graphene-free area, during graphene etching, is an effective roughness effect, stemming from variations of the graphene coverage at length-scales below the coherence length of the electron beam.

The $a_\mathrm{C}$ increase can be interpreted as the buildup of tensile strain induced by the formation of single atom vacancies and interstitials, in agreement with our DFT calculations. The increase goes beyond the strain obtained by DFT for an isolated defect. This is a possible signature of the interaction between defects through the strain fields  they create. In graphene/SiO$_2$/Si, much smaller strains were detected with Raman spectroscopy. About 20~cm$^{-1}$ shifts of the 2D vibration mode were found \cite{MartinsFerreira2010}, corresponding to 0.3\% strains at most under the assumption of biaxial strain \cite{Mohiuddin2009}.

What is the origin for the seven times larger strains measured in graphene/Ir(111)? In graphene/SiO$_2$/Si, the contact between graphene and its substrate is only local due to a relatively high substrate corrugation, at the protruding points of the substrate, and graphene is free-standing elsewhere \cite{Geringer2009}. Interstitials may be trapped without distorting graphene below these free-standing regions. In contrast, the contact between graphene and Ir(111) is rather intimate, characterized by a 3.4~\AA\, average distance \cite{Busse2011}. Strong steric effects are thus expected for interstitials, which should induce noticeable strain fields, in agreement with our DFT calculations [Fig.~\ref{fig3}(b)]. In addition, the loose graphene/SiO$_2$ contact is favorable to nanorippling, rather than bond length compression, in case of interaction between vacancies or interstitials through their strain fields. In graphene/Ir(111), much less freedom exists for nanorippling due to the non vanishing graphene/metal interaction, so that C bond contraction must play an important role.

The broadening of the graphene streaks arises from strain fields, the finite size of structurally coherent domains, and/or a roughening of graphene. The first two effects can be estimated from the inverse of the FWHM of the streaks [Fig.~\ref{fig4}(b)]. The third effect, resulting from the increasing corrugation around the defects or displaced C atoms not escaping the surface, is more difficult to assess. At low doses, finite size effects are not actually relevant, as the coherence length of electrons (typically 10~nm) sets the apparent size of the structurally coherent graphene domains. At larger doses the estimated size falls below this coherence length, indicating that small domains could be present. Strain fields, typically a few 1\%, are also relevant in this regime. All effects are consistent with large defect densities and a tendency to graphene gradual amorphization under irradiation.

The effects of oxygen etching on a full layer of graphene, which is efficient only at elevated temperatures (here 950$^\circ$C), are rather peculiar [Fig.~\ref{fig4}(c)]: first $a_\mathrm{C}$ rapidly increases by only 0.4\%, then very slowly, and graphene streaks only slightly broaden even before vanishing, after a 2.5$\times$10$^{-4}$~mbar$\times$s O$_2$ dose. This dose is close to that corresponding to total removal of graphene, 3$\times$10$^{-4}$~mbar$\times$s, as determined by \textit{in situ} imaging during etching with low-energy electron microscopy in other experimental set-ups (data not shown). The rapid $a_\mathrm{C}$ increase is interpreted as the buildup of strain fields around defects present in low density, vacancies of various sizes, created at the vacancies left in graphene after growth, and/or heptagon-pentagon pairs found at grain boundaries, where local bending and thus reactivity are stronger. Once a critical defect density is reached, O$_2$ etching most likely occurs at the edges of existing vacancies without the need for creating new ones, presumably by graphene decomposition with intercalated oxygen as an intermediate step \cite{Sutter2010,Starodub2010}. Unlike the case of ion bombardment at the highest ion dose explored, qualitative estimates of the size of the structurally coherent graphene domains indicate that they never are below the coherent length of the electrons. The average distance between defects is hence larger than in the case of ion bombardment, and the streaks broadening is ascribed to strains around defects, typically in the range of 1\%.

In conclusion, we have shown that vacancies formed during CVD, ion bombardment, and high temperature oxygen etching, as well as atoms trapped between graphene and its substrate, all induce significant strains in epitaxial graphene. When the system is close to complete amorphization, with defect  separation of less than 1~nm, tensile strains in graphene on Ir(111) reach 2.2\%, a value much larger than in graphene on SiO$_2$. Graphene goes through a series of commensurate phases with its substrate as the vacancy density varies. Besides the strong local perturbations of the properties of graphene due to missing atoms (vacancies), this change of epitaxy is expected to give rise to the changes of the graphene-support interaction. The observed defect-induced strains are, \textit{e.g.}, candidates for engineering the electronic properties of graphene in a straintronics approach, not only around vacancy defect sites \cite{Ugeda2011}, but also where the carbon lattice is not disrupted, in between vacancies and around intercalated atoms or molecules. An analog to a zero-energy Landau level, predicted in nanorippled graphene \cite{Guinea2008}, could indeed develop in the regions where strains vary. This hallmark for the Dirac-fermion-like nature of charge carriers in graphene is for instance expected to strongly enhance graphene's chemical reactivity, and opens the way to strain-promoted  experiments. The effects we observed are expected to take also place during the preparation and processing of a wealth of related systems, for instance the actively investigated graphene/Cu and BN/metal interfaces.

FJ, GR, and JC acknowledge financial support from Agence Nationale de la Recherche through the ANR-2010-BLAN-1019-NMGEM contract and from EU through the NMP3-SL-2010-246073 GRENADA contract. AVK acknowledges the Academy of Finland for the support through projects 218545 and 263416. Valuable help from Y. Cur\'{e} and O. Ulrich is gratefully acknowledged. AKV thanks CSC Finland for generous grants of computer time.


%

\end{document}